\documentclass[10pt,letterpaper,twocolumn]{article} 

\usepackage{ol2}
\usepackage[draft]{hyperref}
\usepackage{amsmath}

\begin{document}

\twocolumn[ 

\title{Bound states in the continuum in $\mathcal{PT}$-symmetric optical lattices}


\author{Stefano Longhi}

\address{Dipartimento di Fisica, Politecnico di Milano and Istituto di Fotonica e Nanotecnologie del Consiglio Nazionale delle Ricerche, Piazza L. da Vinci 32, I-20133 Milano, Italy (stefano.longhi@polimi.it)}

\begin{abstract}
Bound states in the continuum (BIC), i.e. normalizable modes with an energy embedded in the continuous spectrum of scattered states, are shown to exist in certain optical waveguide lattices with $\mathcal{PT}$-symmetric defects. Two distinct types of BIC modes are found: BIC  states that exist in the broken $\mathcal{PT}$ phase,  corresponding to exponentially-localized  modes with either exponentially damped or amplified optical power; and BIC modes with sub-exponential spatial localization that can exist in the unbroken $\mathcal{PT}$ phase as well. The two types of BIC modes at the $\mathcal{PT}$ symmetry breaking point behave rather differently: while in the former case spatial localization is lost and the defect coherently radiates outgoing waves with an optical power that linearly increases with the propagation distance, in the latter case localization is maintained and the optical power increase is quadratic.  
\end{abstract} 



 ] 

\noindent 
In 1929 von Neumann and Wigner suggested \cite{NW} that certain potentials
could support spatially localized states with an energy embedded into the
continuum spectrum of scattered states, so-called bound states in the continuum (BIC). However, BIC states are generally fragile, require specially tailored potentials, and are thus of difficult observation in quantum systems. Photonic structures, allowing robust control of parameters, turned out to be very promising for the observation of BIC states \cite{P1,P2,P3,P4}. In fact, the observation of photonic BIC states have been reported recently by a few groups using waveguide arrays \cite{S1,S2,S3} and photonic crystals \cite{S4}. Most of previous studies on BIC states have been limited to consider Hermitian (i.e. non-dissipative) systems. In recent years, a growing interest has been devoted to study light propagation in so-called $\mathcal{PT}$-symmetric optical structures, which comprise balanced regions of gain and loss. The notion of $\mathcal{PT}$ symmetry, originated within
the context of quantum field theories \cite{Bender}, has
been introduced in optics as a new paradigm to mold the
flow of light \cite{PT1,PT2,PTS,PT3}, enabling to realize unusual and previously
unattainable light propagation features (see, for instance, \cite{PT2,PT3,PT4,PT5,PT6,PT7,PT8} and references therein). In the framework of $\mathcal{PT}$ optics, linear bound states sustained by defects in otherwise periodic lattices have been investigated in a few recent works \cite{B1,B2}. Such states generally fall  in a forbidden lattice band \cite{B1}, like in Hermitian lattices. However, in a recent work \cite{B2} it has been shown, both theoretically and experimentally, that BIC states can arise in the broken $\mathcal{PT}$ phase. Such states, that we will refer to as type-I BIC, are exponentially localized in space and are exponentially damped or amplified along the propagation distance. As the $\mathcal{PT}$ symmetry breaking point is approached from above, type-I BIC modes cease to be localized, however they can coherently radiate outgoing waves \cite{B2} as a signature of a spectral singularity, like in a lasing medium at threshold \cite{PT5,LonghiSS}. So far BIC states in the unbroken $\mathcal{PT}$ phase have not been predicted yet.\\
In this Letter we show that, besides (most common) type-I BIC states found in the broken $\mathcal{PT}$ phase, a novel type of BIC modes, referred to as type-II BIC, can arise in specially-tailored optical lattices. As opposed to type-I BIC, type-II BIC exist above and {\it below} the $\mathcal{PT}$ symmetry breaking, show a lower-than-exponential spatial localization, and behave rather differently than type-I BIC modes at the symmetry breaking point.\\  
\begin{figure}[htb]
\centerline{\includegraphics[width=8.3cm]{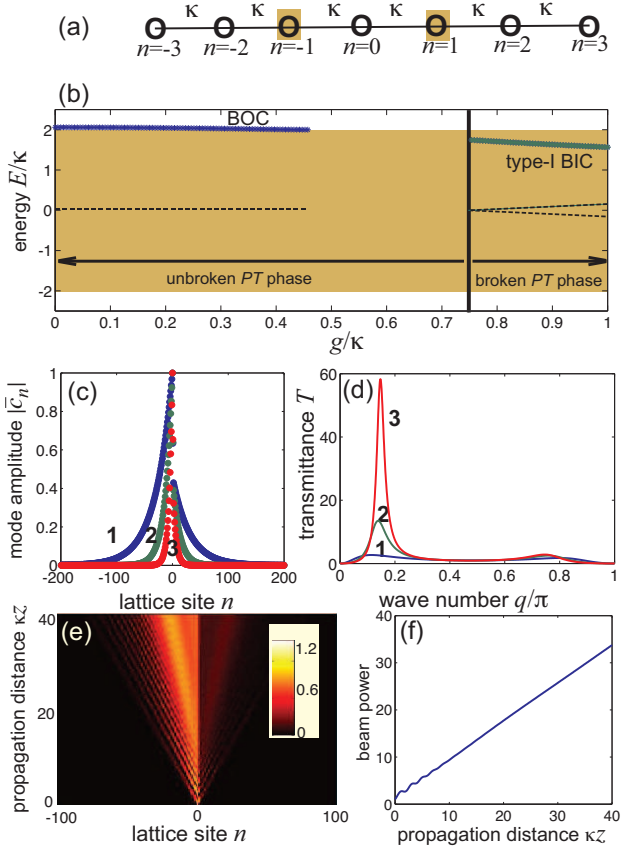}} \caption{
\footnotesize {(Color online)  (a) Schematic of  a homogeneous lattice with two site impurities at waveguides $n= \pm 1$, [Eq.(2) in the text]. (b) Behavior of the energy spectrum (eigenvalues of the matrix $\mathcal{H}$ normalized to the coupling constant $\kappa$) for $\Delta / \kappa=0.3$ and for increasing values of $g/ \kappa$. The shaded region depicts the energy band of scattered states (real eigenvalues). The solid and dashed curves are the real and imaginary parts, respectively, of bound state energies. For small values of $g/ \kappa$ (below $ < \sim 0.457$), there is one BOC state with a real-valued energy which falls just outside the upper band edge. $\mathcal{PT}$ symmetry breaking occurs at $g_{th}/ \kappa \simeq 0.752$, above which two type-I BIC modes with complex conjugate energies emerge. (c) Amplitude distribution ($|\bar{c}_n|$) of the amplified type-I BIC mode for $g/g_{th}=1.08$ (curve 1),  $g/g_{th}=1.2$ (curve 2), and $g/g_{th}=1.5$ (curve 3). (d) Lattice spectral transmittance for $g/g_{th}=0.5$ (curve 1),  $g/g_{th}=0.8$ (curve 2) and $g/g_{th}=0.9$ (curve 3). The resonance peak occurs at $q=q_0 \simeq 0.161 \pi $. (e) Evolution of beam intensity and (f) of normalized total beam power $P(z)/P(0)$ for initial excitation of waveguide $n=0$ at the symmetry breaking transition.}}
\end{figure}
\\
We consider light propagation in an array of evanescently-coupled optical waveguides, which is described by standard coupled mode equations
\begin{equation}
i \frac{dc_n}{dz}=\kappa_nc_{n-1}+\kappa_{n+1}c_{n+1}+V_n c_{n} \equiv \mathcal{H} c_n
\end{equation}
where $c_n=c_n(z)$ is the amplitude of the guided mode at lattice site $n$ and propagation distance $z$, $\kappa_n$ is the coupling constant between waveguides $n$ and $(n-1)$, and $V_n$ is the complex optical potential at site $n$. The potential $V_n$ accounts for both propagation constant shift (the real part of $V_n$) and gain/loss (the imaginary part of $V_n$) in waveguide $n$. $\mathcal{PT}$ invariance requires $\kappa_{-n}=\kappa_{n+1}^*$ and $V_{-n}=V^{*}_n$. In this case, eigenvalues of the matrix system $\mathcal{H}$ are either real-valued or pairs of complex-conjugate numbers: in fact, if $\bar{c}_n$ is an eigenstate of $\mathcal{H}$ with eigenvalue $E$, $\mathcal{H} \bar{c}_n=E \bar{c}_n$, then $\bar{c}_{-n}^*$ is also an eigenstate of $\mathcal{H}$ with eigenvalue $E^*$. The Hamiltonian $\mathcal{H}$ is assumed to depend on a control parameter $g$, which  measures the strength of non-Hermiticity ($g=0$ for the Hermitian case). While at $g=0$ the spectrum of $\mathcal{H}$ is entirely real, as $g$ is increased a threshold value $g=g_{th}$ is generally found, above which pairs of complex conjugate energies emerge. Such a point is usually referred to as the $\mathcal{PT}$ symmetry breaking point.
We consider here the case where a $\mathcal{PT}$-symmetric defect is introduced in an otherwise homogeneous and Hermitian lattice, i.e. we assume that $V_{n} \rightarrow 0$ and $\kappa_n \rightarrow \kappa$ as $n \rightarrow \pm \infty$, where $\kappa$ is the homogeneous real-valued coupling constant far from the defective region.\par Defective type-I BIC modes, such as those studied in Ref.\cite{B2}, are rather common and can be found in very simple lattice models. Less common are type-II BIC states, which exist below, at and above the symmetry breaking point. In this Letter we will consider two examples of waveguide lattices with $\mathcal{PT}$-symmetric defects, which show very different scattering and amplifying properties as a result of the different nature of their BIC states.  \par 
\begin{figure}[htb]
\centerline{\includegraphics[width=8.3cm]{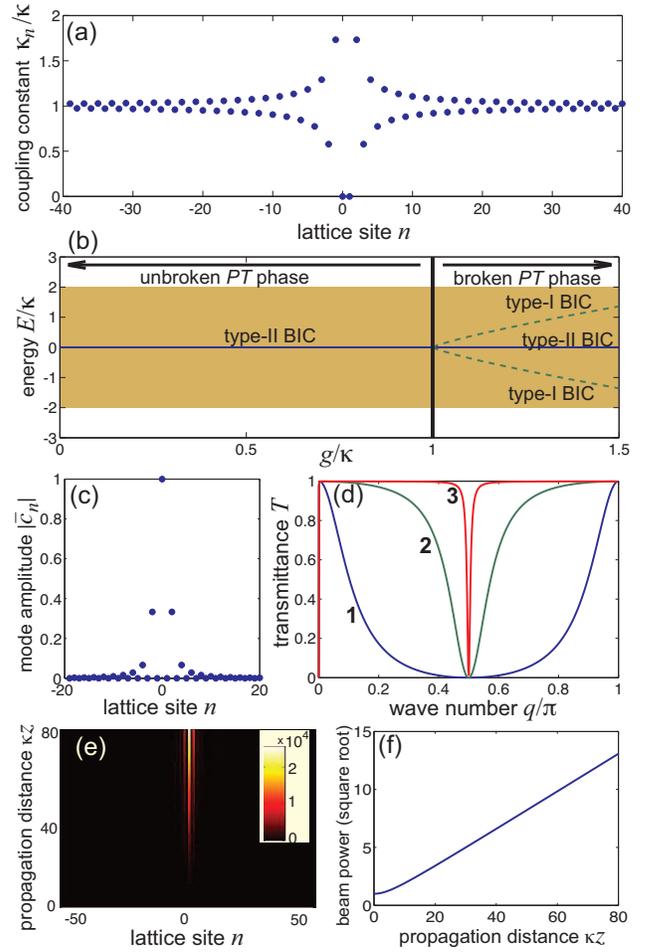}} \caption{
\footnotesize {(Color online)  (a) Behavior of the hopping rates $\kappa_n$ (for $n \neq 0,1$) of the inhomogeneous lattice defined by Eq.(4). (b) Behavior of the energy spectrum of $\mathcal{H}$ for increasing values of $g/ \kappa$. The shaded region depicts the energy band of scattered states (real eigenvalues). The solid and dashed curves are the real and imaginary parts, respectively, of bound state energies. $\mathcal{PT}$ symmetry breaking occurs at $g_{th}/ \kappa =1$, above which two type-I BIC modes with complex conjugate energies emerge. The type-II BIC mode with energy $E=0$ exists for any value of $g$. (c) Amplitude distribution ($|\bar{c}_n|$) of type-II BIC mode at $g=g_{th}$. (d) Lattice spectral transmittance $T(q)$ for $g/g_{th}=0.5$ (curve 1),  $g/g_{th}=0.9$ (curve 2) and $g/g_{th}=0.99$ (curve 3). (e) Evolution of beam intensity and (f) of square root of normalized total beam power (i.e. $\sqrt{P(z)/P(0)}$) for initial excitation of waveguide $n=0$ at the symmetry breaking transition.}}
\end{figure}
 {\it First example. Defective lattice with type-I BIC.} As a first example, let us consider one of the simplest  lattice model with a $\mathcal{PT}$-symmetric defect, corresponding to a uniform lattice with two 
site potential defects \cite{Kottos,uffi}, namely
\begin{equation}
 \kappa_n=\kappa, \;\;\; V_n=\sigma \delta_{n,-1}+\sigma^* \delta_{n,1},
 \end{equation} 
 where $\sigma=\Delta+ig$ is the complex potential, accounting for propagation constant mismatch $\Delta$ and gain/loss $g$ of waveguides at the defective sites $n=\pm 1$ [Fig.1(a)]. For such a simple lattice model bound states with exponential spatial localization, either inside or outside the lattice band $(-2 \kappa, 2 \kappa)$ of Bloch modes, can be found in a closed form by standard methods. For a fixed and relatively small value of the normalized index detuning $\Delta / \kappa$, the spectrum of $\mathcal{H}$ for increasing values of normalized gain $g / \kappa$ shows the typical behavior depicted in Fig.1(b). For low values of $g/ \kappa$, there is a single bound state with a real energy outside the lattice band (i.e. an ordinary bound state outside the continuum, BOC). As $g/ \kappa$ is increased, the spectrum remains real-valued but the BOC disappears, until the $\mathcal{PT}$ symmetry breaking threshold is reached at $g/ \kappa= g_{th}/\kappa$. Above the symmetry breaking threshold, a couple of bound states with complex conjugate energies, embedded in the lattice band, emanate at some real energy $E_0=2 \kappa \cos q_0$, with a localization length that diverges as $g$ approaches $g_{th}$ from above, see Fig.1(c). These are precisely type-I BIC modes, previously observed in Ref.\cite{B2} in a defective mesh lattice. For an initial excitation of the lattice at site $n=0$, for $g>g_{th}$ light trapping is observed near $n=0$ owing to the existence of  the bound states, and the optical power exponential increases with the propagation distance as a result of the broken $\mathcal{PT}$ phase (complex energies). Interestingly, at the $\mathcal{PT}$ symmetry breaking point light is no more trapped near  $n=0$ because of the disappearance of type-I BIC modes, however undamped coherent emission of outgoing waves from the defective region is observed; see Fig.1(e). Note that the local intensity $|c_n(z)|^2$ remains bounded while the optical beam power $P(z)=\sum_n |c_n(z)|^2$ increases linearly with $z$ as a result of outgoing wave emission; see Fig.1(f). Such a result, which was observed in Ref.\cite{B2}, is the typical behavior of $\mathcal{PT}$ symmetry breaking associated to the appearance of a spectral singularity \cite{LonghiSS,Mostafazadeh}. This can be seen by looking at the behavior of the spectral transmission and reflection coefficients of the lattice across the defective region. The spectral transmission coefficient $t$ for a Bloch wave with wave number $q$ can be readily calculated by standard methods and reads 
 \begin{equation}
 t(q)=\frac{i \sin q \exp(2iq)}{\Delta+i \sin q-|\sigma|^2 \cos q \exp(2iq)}.
 \end{equation}
A typical behavior of the spectral transmittance $T(q)=|t(q)|^2$ for a few values of $g/g_{th}<1$ is shown in Fig.1(d). Note that a strong resonance peak near $q=q_0$ is found, which narrows and diverges as the symmetry breaking point is approached. This is a clear signature of a spectral singularity \cite{Mostafazadeh} and it is a typical scenario of a lasing instability that arises in an active optical cavity \cite{LonghiSS}. 
\par
 {\it Second example. Defective lattice with type-II BIC.} 
Type-II BIC modes, which exist in the {\it unbroken} $\mathcal{PT}$ phase, have not been discussed so far. Like for Hermitian lattices \cite{P4,S2}, such states are expected to show a lower than exponential localization and to arise in specially-engineered lattices, i.e. they are much less common than type-I BIC. An example of a $\mathcal{PT}$ symmetric lattice that sustains type-II BIC is provided below. We assume a lattice with vanishing potential, i.e. $V_n=0$, and with inhomogeneous coupling constants given by
\begin{equation} 
\frac{\kappa_n} { \kappa} =\left\{ 
\begin{array}{lll}
\sqrt{(n+1)/(n-1)} & n \; {\rm even} \;, \; n \neq 0 & \kappa_0=-ig \;\;\;\;\\
\sqrt{(n-2)/n} & n \; {\rm odd} \;, \; n \neq 1 & \kappa_{1}=ig  \;\;\;\;
\end{array}
\right.
\end{equation}
where $g>0$ is a real-valued parameter. The behavior of $\kappa_n$ is shown in Fig.2(a). Note that, since $\kappa_n / \kappa \rightarrow 1$ as $n \rightarrow \pm \infty$, the lattice is asymptotically homogeneous. It also satisfies the $\mathcal{PT}$ symmetry requirement  $\kappa_{-n}=\kappa_{n+1}^*$. The non-Hermitian nature of the lattice arises, in this case, by the imaginary value of the coupling constants $\kappa_0$ and $\kappa_1$. In coupled optical waveguides, an effective imaginary coupling constant can be realized by suitable longitudinal modulation of index and gain/loss terms, as discussed in Ref.\cite{LonghiPRA10}. The inhomogeneous coupling constants $\kappa_n$  for $n \neq 0,1$ can be readily obtained by judicious waveguide spacing, as demonstrated in Ref.\cite{S2}. Numerical analysis of the eigenvalues of the matrix $\mathcal{H}$ shows that the energy spectrum is real-valued for $g \leq g_{th}=\kappa$, i.e. $\mathcal{PT}$ symmetry breaking occurs at $g_{th}=\kappa$. In the unbroken $\mathcal{PT}$ phase ($g < g_{th}$), the energy spectrum comprises, in addition to the continuous spectrum $(-2 \kappa, 2 \kappa)$ of scattered states of the asymptotic homogeneous lattice, one BIC mode with algebraic localization at the energy $E=0$, i.e. a type-II BIC [see Fig.2(b)]. It can be readily checked from Eqs.(1) and (4) that the BIC mode  is given by (apart from a normalization factor):
\begin{equation} 
\bar{c}_n =\left\{ 
\begin{array}{ll}
0 & n \; {\rm odd}  \;\;\;\;\\
\kappa /g & n=0 \\
\frac{n}{|n|} \frac{i^{n+1}}{ \sqrt{n^2-1}} & n \; {\rm even} ,\; n \neq 0 \;\;\;\;
\end{array}
\right.
\end{equation}
In the broken $\mathcal{PT}$ phase ($g> g_{th}$), two additional type-I BIC modes, which complex conjugate energies and vanishing real part that emanate from $E=0$, are found [see Fig.2(b)]. Such additional BIC states have the same properties as those found in the previous example, and are thus not discussed here anymore.  Note that type-II BIC mode [Eq.(5)] does exist below, at and above the symmetry breaking point, and that by varying the $g$ parameter the amplitude of the $n=0$ site solely is modified. A typical behavior of the amplitude distribution $|\bar{c}_n|$ of type-II BIC mode is shown in Fig.2(c). At $g=g_{th}$, it can be shown that the lattice (4) belongs to a rather general class of non-Hermitian lattices that can be synthesized from the homogeneous lattice $\kappa_n=\kappa$, $V_n=0$ by a double discrete Darboux (supersymmetric) transformation; mathematical details are rather cumbersome and will be given elsewhere \cite{unp}. Here we mention two important results of the analysis, which show that the scattering and amplifying properties of lattice (4) at the symmetry breaking point are very distinct than those of lattice (2):\\ (i) The supersymmetric transformation ensures that the spectral transmission across the defective region, at any energy $E \neq 0$ (i.e. Bloch wave number $q \neq \pi/2$), approaches unity as $g \rightarrow g_{th}^{-}$ [$t(q)=1$ for $ q \neq \pi/2$], i.e. the lattice becomes reflectionless as the symmetry breaking point is approached. This is clearly shown in Fig.2(d), which shows the numerically-computed spectral transmittance $T(q)$ for a few values of $g/g_{th}<1$. Note that $T(q)$ shows a dip (rather than a resonance, as in  the previous example) near $q=\pi/2$; as the symmetry breaking point is attained, the dip narrower and one obtains $T(q)=1$ for $q \neq \pi/2$.\\ (ii) At the symmetry breaking point, the energy $E=0$ is an exceptional point in the continuous spectrum \cite{Andrianov}. Exceptional points in the continuous spectrum (not to be confused with more common exceptional points in finite-dimensional non-Hermitian systems and with spectral singularities in non-Hermitian systems with purely continuous spectrum) and their mathematical properties have been discussed for the continuous Schr\"{o}dinger equation by Andrianov and Sokolov in Ref.\cite{Andrianov}. The circumstance that $E=0$ is an exceptional point in the continuum means that there exists an associated function $f_n$ to type-II BIC mode $\bar{c}_n$, which is limited at $n \rightarrow \pm \infty$ and given by $f_n=-(i/2) \sin ( n \pi /2)$, satisfying the equation $\mathcal{H}f_n=\bar{c}_n$ \cite{Andrianov}. Hence, at the symmetry breaking point Eq.(1) is satisfied not only by the stationary solution $c_n(z)= \bar{c}_n$, but also by any (non-stationary) solution
\begin{equation}
c_n(z)=(1-i \epsilon z) \bar{c}_n+\epsilon f_n
\end{equation}
for any arbitrary value of $\epsilon$. Equation (6) has an important physical implication: it shows that at the symmetry breaking point a small perturbation shaped like the associated function $f_n$ leads to a secular growth with propagation distance of the type-II BIC state $\bar{c}_n$, with an optical power $P(z)$ that increases {\it quadratically} (rather than linearly) with $z$. This is clearly shown in Figs.2(e) and (f), which depict the evolution along the array of the beam intensity and total beam  power for excitation of waveguide $n=0$ at the input plane.\par
In conclusion, we have shown that different types of BIC states can exist in optical lattices with $\mathcal{PT}$ symmetric defects, and that their nature can deeply modify  the scattering and amplifying properties of the lattice at the symmetry breaking point. In particular, we have introduced a defective waveguide lattice which close to the symmetry breaking threshold is reflectionless and shows a rather uncommon quadratic amplification law for the optical power in the guided (BIC) mode, owing to the appearance of an exceptional point in the continuum.\\

\par
The author acknowledges hospitality at the IFISC
(CSIC-UIB), Palma de Mallorca.

\newpage

\newpage

\footnotesize {\bf References with full titles}\\
\\
\noindent
1. J. von Neumann and E. Wigner, {\it \"Uber merkw\"urdige diskrete Eigenwerte}, Z. Phys. {\bf 30}, 465 (1929).\\
2. E.N. Bulgakov and A.F. Sadreev, {\it Bound states in the continuum in photonic waveguides inspired by defects}, Phys. Rev. B {\bf 78},  075105 (2008).\\
3. D. C. Marinica and A. G. Borisov, {\it Bound States in the Continuum in Photonics}, Phys. Rev. Lett. {\bf 100}, 183902 (2008).\\
4. N. Prodanovic, V. Milanovic, and J. Radovanovic, {\it Photonic crystals with bound states in continuum and their realization by an advanced digital grading method}, J. Phys. A {\bf 42}, 415304 (2009).\\
5. M. I. Molina, A. E. Miroshnichenko, and Y. S. Kivshar, {\it Surface Bound States in the Continuum}, Phys. Rev. Lett. {\bf 108}, 070401 (2012).\\
6. Y. Plotnik, O. Peleg, F. Dreisow, M. Heinrich, S. Nolte, A. Szameit, and M. Segev, {\it Experimental Observation of Optical Bound States in the Continuum}, Phys. Rev. Lett. {\bf 107}, 183901 (2011).\\
7. G. Corrielli, G. Della Valle, A. Crespi, R. Osellame, and S. Longhi, {\it Observation of Surface States with Algebraic Localization}, Phys. Rev. Lett. {\bf 111}, 220403 (2013).\\
8. S. Weimann, Y. Xu, R. Keil, A. E. Miroshnichenko, S. Nolte, A. A. Sukhorukov, A. Szameit, and Y. S. Kivshar, {\it Compact surface Fano states embedded in the continuum of waveguide arrays}, Phys. Rev. Lett. {\bf 111}, 240403 (2013).\\
9. C. W. Hsu, B. Zhen,	J. Lee, S.-L. Chua, S.G. Johnson, J.D. Joannopoulos, and M. Soljacic, {\it Observation of trapped light within the radiation continuum}, Nature {\bf 499}, 188 (2013).\\
10. C.M. Bender, {\it Making sense of non-Hermitian Hamiltonians}, Rep. Prog. Phys. {\bf 70}, 947 (2007).\\
11. R. El-Ganainy, K. G. Makris, D. N. Christodoulides, and Z.H. Musslimani, {\it Theory of coupled optical $\mathcal{PT}$ symmetric structures}, Opt. Lett. {\bf 32}, 2632 (2007).\\
12. K. G. Makris, R. El-Ganainy, D. N. Christodoulides, and Z. H. Musslimani, {\it Beam Dynamics in $\mathcal{PT}$-Symmetric Optical Lattices}, Phys. Rev. Lett. {\bf 100}, 103904 (2008).\\
13. C. E. R\"{u}ter, K.G. Makris, R. El-Ganainy, D.N. Christodoulides, M. Segev, and D. Kip, {\it Observation of parityÐtime symmetry in optics}, Nature Phys. {\bf 6}, 192 (2010).\\
14. A. Regensburger, C. Bersch, M.-A. Miri, G. Onishchukov, D.N. Christodoulides, and	U. Peschel, {\it Parity-time synthetic photonic lattices}, Nature {\bf 488}, 167 (2012).\\
15. S. Longhi, {\it Bloch Oscillations in Complex Crystals with $\mathcal{PT}$ Symmetry}, Phys. Rev. Lett. {\bf 103}, 123601 (2009).\\ 
16. S. Longhi, {\it $\mathcal{PT}$ symmetric laser-absorber}, Phys. Rev. A {\bf 82}, 031801 (2010).\\
17. Z. Lin, H. Ramezani, T. Eichelkraut, T. Kottos, H. Cao, and D.N. Christodoulides, {\it Unidirectional Invisibility Induced by $\mathcal{PT}$-Symmetric Periodic Structures},
Phys. Rev. Lett. {\bf 106}, 213901 (2011). \\
18. L. Feng, Y.-L. Xu, W.S. Fegadolli, M.-H. Lu, J.E.B. Oliveira, V.R. Almeida,	 Y.-F. Chen, and A. Scherer, {\it Experimental demonstration of a unidirectional reflectionless parity-time metamaterial at optical frequencies},     Nature Mat. {\bf 12}, 108 (2013).\\
19. T. Eichelkraut, R. Heilmann, S. Weimann,	 S. St\"{u}tzer, F. Dreisow, D.N. Christodoulides, S. Nolte, and A. Szameit, "Mobility transition from ballistic to diffusive transport in non-Hermitian lattices",  Nature Commun. {\bf 4}, 2533 (2013).\\
20. K. Zhou, Z. Guo, J. Wang, and S. Liu, {\it Defect modes in defective parity-time symmetric
periodic complex potentials}, Opt. Lett. {\bf 35}, 2928 (2010).\\
21. A. Regensburger, M.-A. Miri, C. Bersch, J. N\"{a}ger, G. Onishchukov, D.N. Christodoulides, and U. Peschel, {\it Observation of Defect States in $\mathcal{PT}$-Symmetric Optical Lattices}, Phys. Rev. Lett. {\bf 110}, 223902 (2013).\\ 
22.  S. Longhi, {\it Optical Realization of Relativistic Non-Hermitian Quantum Mechanics}, Phys. Rev. Lett. Phys. {\bf 105}, 013903 (2010).\\
23.  O. Bendix, R. Fleischmann, T. Kottos, and B. Shapiro, {\it Exponentially Fragile $\mathcal{PT}$ Symmetry in Lattices with Localized Eigenmodes}, 
Phys. Rev. Lett. {\bf 103}, 030402 (2009).\\
24. Y.N. Joglekar, D. Scott, and M. Babbey, {\it Robust and fragile $\mathcal{PT}$-symmetric phases in a tight-binding chain}, Phys. Rev. A {\bf 82}, 030103(R) (2010).\\
25. A. Mostafazadeh, {\it Spectral Singularities of Complex Scattering Potentials and Infinite Reflection
and Transmission Coefficients at Real Energies}, Phys. Rev. Lett. {\bf 102}, 220402 (2009).\\
26. S. Longhi, {\it Invisibility in non-Hermitian tight-binding lattices}, Phys. Rev. A {\bf 82}, 032111 (2010).\\
27. S. Longhi and G. Della Valle, {\it Optical lattices with exceptional points in the continuum} (unpublished).\\
28. A.A. Andrianov and A.V. Sokolov, {\it Resolutions of Identity for Some Non-Hermitian
Hamiltonians. I. Exceptional Point in Continuous Spectrum}, SIGMA {\bf 7}, 111 (2011).

\end{document}